\documentclass[%
 aip,
 amsmath,amssymb,
 reprint,%
]{revtex4-2}

\usepackage{graphicx}  
\usepackage{dcolumn}   
\usepackage{bm}        
\usepackage{upgreek}   

\usepackage[utf8]{inputenc}
\usepackage[T1]{fontenc}
\usepackage{mathptmx}
\usepackage{etoolbox}

\newcommand{\al}{\alpha}
\newcommand{\be}{\beta}
\newcommand{\ka}{\kappa}
\newcommand{\om}{\omega}

\newcommand{\myten}[1]{\overline{\overline{#1}}}  

\newcommand{\myvec}[1]{\mathbf{#1}}              

\newcommand{\myhat}[1]{\hat{\mathbf{#1}}}        
\newcommand{\myhatGr}[1]{\hat{\bm{#1}}}          

\newcommand{\vecA}{\myvec A}
\newcommand{\vecB}{\myvec B}
\newcommand{\vecE}{\myvec E}
\newcommand{\vecF}{\myvec F}
\newcommand{\vecJ}{\myvec J}
\newcommand{\vecR}{\myvec R}

\newcommand{\veck}{\myvec k}
\newcommand{\vecq}{\myvec q}
\newcommand{\vecu}{\myvec u}

\newcommand{\hatb}{\myhat b}
\newcommand{\hatphi}{\myhatGr\upphi}

\newcommand{\p}{\partial}
\newcommand{\rp}{\right)}
\newcommand{\lp}{\left(}

\makeatletter
\def\@email#1#2{%
 \endgroup
 \patchcmd{\titleblock@produce}
  {\frontmatter@RRAPformat}
  {\frontmatter@RRAPformat{\produce@RRAP{*#1\href{mailto:#2}{#2}}}\frontmatter@RRAPformat}
  {}{}
}%
\makeatother
\begin{document}

\preprint{AIP/123-QED}

\title[Sample title]{Design and Modeling of Indirectly-Driven Magnetized Implosions on the NIF}
\author{D. J. Strozzi}
 \email{strozzi2@llnl.gov.}
\author{H. Sio}
\author{G. B. Zimmerman}
\author{J. D. Moody}
\author{C. R. Weber}
\author{B. Z. Djordjevi\'c}
\author{C. A. Walsh}
\author{B. A. Hammel}
\author{B. B. Pollock}
\author{A. Povilus}
\affiliation{Lawrence Livermore National Laboratory, Livermore, CA 94550}

\author{J. P. Chittenden}
\author{S. O'Neill}
\affiliation{Centre for Inertial Fusion Studies, The Blackett Laboratory, Imperial College, London SW7 2AZ, United Kingdom}

\date{\today}

\begin{abstract}
  The use of magnetic fields to improve the performance of hohlraum-driven implosions on the National Ignition Facility (NIF) is discussed.  The focus is on magnetically insulated inertial confinement fusion (ICF), where the primary field effect is to reduce electron-thermal and alpha-particle loss from the compressed hotspot (magnetic pressure is of secondary importance).  We summarize the requirements to achieve this state.  The design of recent NIF magnetized hohlraum experiments is presented.  These are close to earlier shots in the three-shock, high-adiabat (BigFoot) campaign, subject to the constraints that magnetized NIF targets must be fielded at room-temperature, and use $\lesssim 1$ MJ of laser energy to avoid risk of optics damage from stimulated Brillouin scattering.  We present results from the original magnetized hohlraum platform, as well as a later variant which gives higher hotspot temperature. In both platforms, imposed fields (at the capsule center) of up to 28 T increase the fusion yield and hotspot temperature. Integrated radiation-magneto-hydrodynamic (rad-MHD) modeling with the Lasnex code of these shots is shown, where laser power multipliers and a saturation clamp on cross-beam energy transfer (CBET) are developed to match the time of peak capsule emission and the $P_2$ Legendre moment of the hotspot x-ray image.  The resulting fusion yield and ion temperature agree decently with the measured relative effects of the field, although the absolute simulated yields are higher than the data by $2.0-2.7\times$.  The tuned parameters and yield discrepancy are comparable for experiments with and without an imposed field, indicating the model adequately captures the field effects.  Self-generated and imposed fields are added sequentially to simulations of one BigFoot NIF shot to understand how they alter target dynamics.
\end{abstract}

\maketitle

\section{\label{sec:intro}Introduction}

Magnetized inertial confinement fusion (ICF) is an old idea \cite{linhart-liners-nf-1962, lindemuth-magicf-nf-1983, jones-mead-magicf-nf-1986} that has received significant interest in recent years.  The basic concept is to reduce the loss of energy (carried by electrons and fusion-generated charged particles) from the implosion hotspot, and thereby relax the requirements for ignition and energy gain. This thermal insulation and resulting improved fusion performance have been demonstrated in the three major approaches to ICF: laser direct drive, magnetic direct drive, and indirect (x-ray) drive.  Laser-driven cylindrical and spherical implosions at the OMEGA laser facility have compressed externally-imposed B fields to large values \cite{knauer-bfield-pop-2010, chang-sphere-prl-2011, hohenberger-b-omega-pop-2012} and demonstrated increased fusion yield and ion temperature $T_i$. The magnetically-driven MagLIF concept uses the pulsed-power Z Machine at Sandia National Laboratories to implode a cylindrical metal liner that contains fusion fuel with an imposed axial magnetic field and preheated by a laser \cite{slutz-maglif-prl-2012, gomez-maglif-prl-2014}.  The imposed field is needed to confine DT fusion alpha particles, and has been shown experimentally to improve performance.  An imposed magnetic field can also improve the fast ignition concept \cite{tabak-fi-pop-1994} by guiding relativistic electrons to the hotspot and mitigate their angular divergence. This was proposed in a modeling study \cite{strozzi-fastig-pop-2012} and has recently been explored in experiments with kilotesla laser-driven fields \cite{sakata-magFI-natcomm-2018}. 

This paper focuses on magnetized indirect-drive ICF, which has been researched at Lawrence Livermore National Lab (LLNL) for over a decade \cite{perkins-magicf-pop-2013, moody-magicf-jfe-2022}.  Recent experiments at the National Ignition Facility (NIF) \cite{moses-nif-fuscitech-2016} have studied hohlraum-driven implosions of gas-filled capsules, with imposed initial axial magnetic field up to 28 tesla \cite{moody-maghohl-prl-2022, sio-maghohl-pop-2023}. These have demonstrated increased fusion yield and $T_i$, in line with radiation-magneto-hydrodynamic (rad-MHD) modeling.  We continue to build on this work, with the goal of using magnetic fields to enhance high-yield designs in the ignition regime.

The role of magnetic fields in fusion devices can be understood by considering two dimensionless measures: the electron Hall parameter $H_e \equiv \om_{ce}\tau_{ei}$ ($\om_{ce}\equiv eB/m_e$ is the electron cyclotron frequency and $\tau_{ei}$ the electron-ion collision time), and $\beta \equiv 2\mu_0p/B^2$ ($p$ is the thermal pressure).  For all systems currently under study, the minimum requirement to be magnetized is $H_e > 1$.  This means the electron thermal conductivity $\ka$ perpendicular to $\vecB$ is significantly reduced.  Within this class, systems can be divided based on $\beta$.  For $\beta \gg 1$, magnetic pressure has a small direct effect, and we call this regime ``magnetically-insulated'' ICF.  Magnetized ICF schemes for NIF and other laser facilities fall in this domain.  $\beta \sim 1$ is the ``magneto-inertial fusion'' regime, where magnetic and thermal pressures are comparable \cite{lindemuth-mif-pop-2017}.  This includes the magnetic compression stage of pulsed-power driven MagLIF, though the imploded fuel has $\beta \gg 1$ (so the laser-driven ``mini-MagLIF'' scheme at OMEGA \cite{barnak-minimaglif-pop-2017} is magnetically-insulated ICF).  Other interesting MIF approaches include the sheared-flow stabilized Z-pinch \cite{zhang-shearflowstabZ-prl-2019} and the dense plasma focus \cite{krishnan-dps-ieeetps-2012}.  $\beta \ll 1$, where magnetic pressure dominates the confinement, is the regime of classic magnetic-fusion systems like tokamaks and stellarators.

We now estimate the magnetic field requirements for magnetizing a typical NIF hohlraum-driven hotspot.  High-performing capsules with a cryogenic DT ice layer have a measured convergence ratio $CR=$ (initial capsule radius) / (stagnated hotspot radius) $\sim 30$.  We consider an initial axial $B$ field of a uniform 40 T in the capsule, and make the optimistic assumption that magnetic flux is conserved in the implosion according to the frozen-in law of ideal MHD.  The field on the equator is then enhanced by $CR^2 \sim 900$, giving $B \sim$ 36 kT. A marginally igniting hotspot has density $\sim$ 100 mg/cm$^3$ and temperature $\sim$ 6 keV, which gives Hall parameter $H\sim$ 9.5.  To estimate the effect on $\ka$, for a purely axial field the effective $\ka/\ka_{||} \sim 1/3 + (2/3)\ka_\perp/\ka_{||}$ with $\ka_{||}$ the unmagnetized, axial value.  $\ka$ is reduced in the two perpendicular directions.  For $Z=1$ and $H=9.5$, the Epperlein and Haines \cite{epperlein-xport-pof-1986} fits for magnetized $\ka$ give almost full suppression: $\kappa_\perp/\kappa_{||} = 0.012$. Even $H=1$ gives significant reduction: $\kappa_\perp/\kappa_{||} = 0.32$. These parameters thus give substantial margin for processes that reduce the benefits of magnetization.  These include flux loss due to effects like resistive diffusion and the Nernst effect, or non-axial imploded field geometries.  From this simple estimate and detailed rad-MHD modeling, the LLNL program aims to impose axial fields up to 100 T on layered cryogenic capsules driven by NIF hohlraums.

This paper is organized as follows. Section 2 describes the constraints on, design of and results from magnetized NIF hohlraum-driven gas-filled capsule experiments.  Rad-MHD modeling of these shots with the Lasnex code and Lasnex Hohlraum Template (LHT) common model are presented in Sec.\ 3.  Sec.\ 4 contains a modeling study of MHD and imposed-field effects on one NIF shot from the BigFoot campaign, upon which the magnetized platform is based.  Conclusions and future work toward magnetized ignition on NIF are given in Sec.\ 5.  Two appendices give details on the Lasnex LHT model and Lasnex's MHD package.

\section{Magnetized Hohlraum-Driven Implosions of Gas-Filled Capsules on NIF}

This section presents the rationale behind and results of NIF experiments with hohlraum-driven gas-filled capsules, with and without imposed magnetic field.  This work has demonstrated increased hotspot ion temperature and yield with a field, and has been partially reported elsewhere \cite{moody-maghohl-prl-2022, sio-maghohl-pop-2023}.  The new aspects here are a detailed discussion of the ``WarmMag'' platform's design rationale and early results from recent experiments with increased hotspot temperature (with or without field).

\subsection{``WarmMag'' Magnetized Room-Temperature NIF platform constraints}

Our ``WarmMag'' platform is designed to demonstrate magnetic insulation of an indirect-drive ICF hotspot, subject to various practical constraints.  The platform has three designs: the initial low-T design with prolate implosion shape, the final low-T design with round shape when magnetized, and the high-T design with increased capsule hotspot temperature. We magnetize a NIF hohlraum by wrapping it with an external coil and running current through it via a pulsed-power system.  The most salient present constraint is that one can only magnetize room-temperature (293 K) NIF experiments.  A room temperature pulsed power system consisting of a 4 $\mu$F capacitor and a spark gap trigger system has been in use for a variety of NIF target experiments since 2016.  Pre-magnetizing cryogenic DT-layered implosion targets requires integration of a pulsed power system with a cryogenic target positioner.  NIF has designed this system but the completion date for construction is currently under evaluation.  Room-temperature hohlraum implosions are fielded somewhat routinely on NIF \cite{ralph-warmhohl-pop-2016}.  They are limited to gas-filled capsules or ``symcaps,'' with no ice layers and no keyhole shock-timing experiments (these require a re-entrant gold cone filled with liquid D$_2$ inserted into the capsule).

There are limits on the room-temperature pressures and compositions of capsule and hohlraum gas fills.  Our diamond or high density carbon (HDC) capsule can hold very high pressures of H and He isotopes without permeation, so this does not affect our gas choice.  The hohlraum gas fill is limited by not bursting the laser entrance hole (LEH) windows, for which the probability gradually increases with pressure.  To minimize the likelihood of bursting, we typically use $\sim 1\ \mu$m thick windows vs.\ $\sim 1/2\ \mu$m for cryo experiments. We also follow the common NIF practice of using neopentane (C$_5$H$_{12}$) for room-temperature hohlraum fills, with the same fully-ionized electron density as the He in the analogous cryo design \cite{ralph-warmhohl-pop-2016}.  This corresponds to much lower initial fill pressure.

Another constraint is having the field permeate the hohlraum and capsule before the laser fires, without significantly perturbing either.  The field must soak through the high-Z hohlraum with acceptably small $J \times B$ force, Joule heating, and temporal delay due to resistive diffusion \cite{moody-maghohl-pop-2020}.  All these effects increase with hohlraum conductivity. The most limiting for our current gas-filled experiments is the $J \times B$ force.  We require the resulting radial wall motion to be less than $50\ \mu$m, which is comparable to NIF laser beam pointing error.  Pure gold hohlraums are too conductive to meet this, though pure uranium hohlraums might be acceptable.  Future experiments with ignition-quality cryogenic DT ice layers have more stringent constraints which require lower conductivity.  For instance, the hohlraum and coil must allow adequate thermal control and spatial uniformity, and radiation from a Joule-heated hohlraum must not disturb the ice layer.  Uranium is probably not adequate for layered targets with 60-70 T imposed field, so instead we use a novel, resistive hohlraum alloy AuTa$_4$ \cite{bayu-aji-AuTa-jpd-2021, shin-auta-jap-2024}.  Our HDC capsule is non-conductive so field soak-through poses no challenges, though it may for a conductive capsule made of e.g.\ beryllium.

The final constraint due to magnetization is the risk of laser optics damage due to SBS \cite{kirkwood-sbs-dpp-2014}.  Magnetic fields introduce risk like any new target, and laser energy and power must be increased in safe increments.  The new risk with fields is that the pulsed-power system could prefire before the laser.  The current melts the coils and induces $J\times B$ motion, all of which can leave the target in an unknown state (with unknown SBS risk) when the laser fires.  Pulser prefire has not been observed on any NIF shot to date, but its risk is unacceptably high.  Every magnetized NIF shot must therefore pass a stringent SBS risk rule, which can only be relaxed when the prefire risk is made small enough.  The SBS rule is a limit on the ``equivalent fluence'' of laser power vs.\ time, which is a complicated functional somewhat between total energy and peak power \cite{carr-damage-apl-2007}.  This usually limits hohlraum pulses to $\sim$ 5 kJ/beam (or 1 MJ full NIF). A low-power 1 TW/beam ``caboose'' at the end of the pulse for a standard hohlraum is considered to not add risk.  The planned cryo pulser will include prefire mitigation and remove the SBS limit.

\begin{figure}
  \centering
  \includegraphics[width=3in]{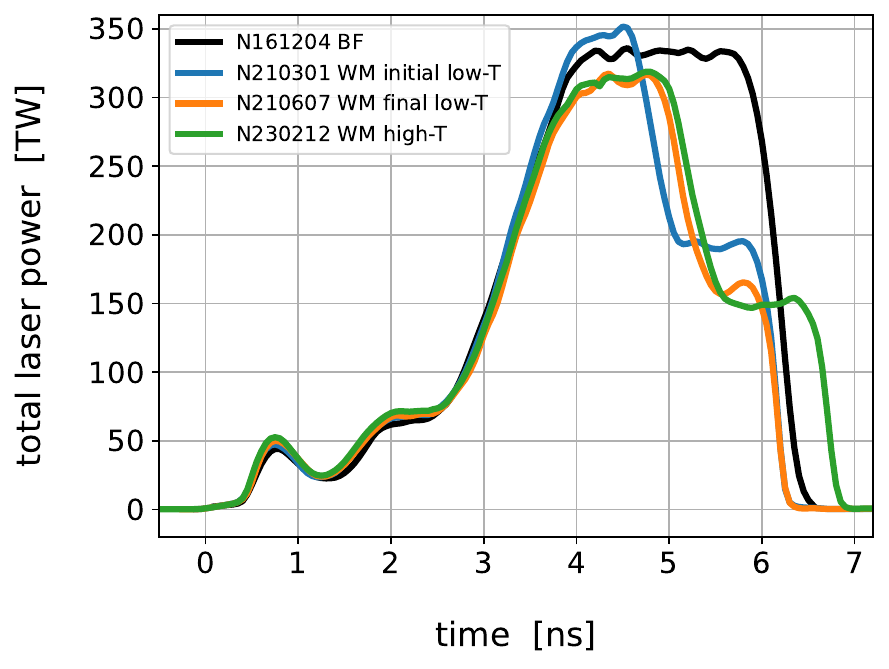}
  \includegraphics[width=3in]{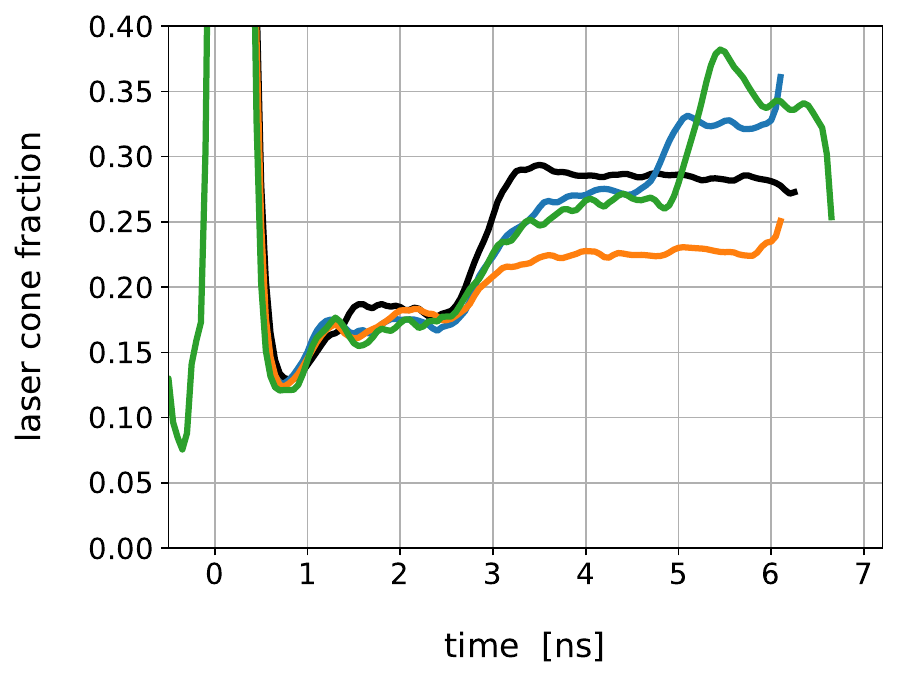}
  \caption{Total laser power (top) and cone fraction (bottom: inner power / total power) vs.\ time for four gas-filled capsule NIF experiments.  BigFoot in black, WarmMag (WM) initial low-T in blue, WM low-T in orange, WM high-T in green.  Cone fraction is only plotted until laser power falls below 50 TW.}
  \label{fig:gascap-pulses}
\end{figure}

\subsection{``WarmMag'' initial low-T platform: BigFoot analog, prolate hotspot}

The room-temperature magnetized hohlraum project aims to understand magnetization with $\lesssim$ 5 shots per year which each use $\lesssim$ 1 MJ of laser energy.  We therefore start from an existing NIF platform that has both high performance and robustness, meaning shots are likely to be reproducible and not depart much from expectations.  The BigFoot platform stands out in this regard \cite{thomas-bigfoot-pop-2020, baker-bigfoot-prl-2018}.  It was deliberately designed for robustness, with a high implosion adiabat and high density HDC capsule.  Both of these lead to short laser pulses, which require a relatively low density hohlraum gas fill of 0.3 mg/cm$^3$ of He to adequately limit inward hohlraum wall motion.  This has been demonstrated to produce much less backscatter than longer-pulse, higher-fill platforms like the earlier LowFoot and HighFoot.  BigFoot achieved round implosions without using wavelength shifts $\Delta \lambda$ between the NIF lasers.  This is done to control cross-beam energy transfer (CBET)\cite{kruer-cbet-pop-1996, michel-xbeam-prl-2009}, a nonlinear laser-plasma interaction that transfers power to the longer-wavelength laser in the plasma flow frame.  $\Delta \lambda=0$ for all BigFoot and WarmMag shots.  BigFoot has also given the highest fusion yield (29.9 kJ) of any NIF platform with $\leq$ 1.3 MJ of laser energy, on NIF shot N190721-2 \cite{baker-alpha-heating-pre-2023}.  The record yield NIF shots use the Hybrid E platform \cite{abu-Shawareb-breakeven-prl-2024} and have given much more $\sim 60\times$ yield than BigFoot, but this platform uses at least 1.8 MJ of laser which our SBS rule precludes.  Besides that, new ideas are more affordably tested with less laser energy and power, since optics damage increases rapidly with both. N190721-2 is an appealing BigFoot layered target for a baseline magnetized design.  This was used in the alpha-heating campaign \cite{baker-alpha-heating-pre-2023} and gave an inferred yield amplification due to alpha heating of $\sim 2.5\times$.  Adding a initial B field $B_{z0} \geq 20$ T (quoted at the capsule center) should measurably enhance the yield due to both reduced electron thermal loss and enhanced alpha heating.

\begin{figure*}    
  \centering
  \includegraphics[width=6in]{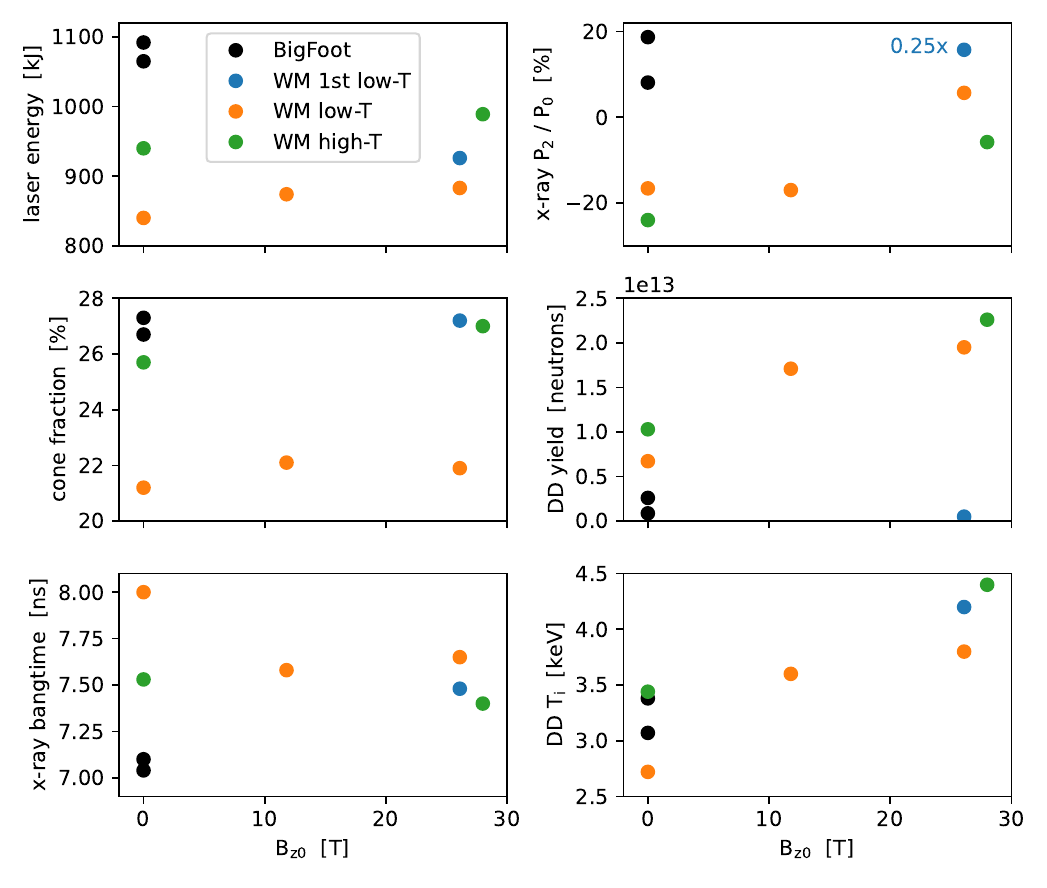}  
  \caption{Data for NIF shots in Table 1, same color scheme as Fog.\ \ref{fig:gascap-pulses}.  Cone fraction is (inner-beam energy) / (total laser energy).}
  \label{fig:nifshots-expt}
\end{figure*}

\begin{figure*}    
  \centering
  \includegraphics[width=6in]{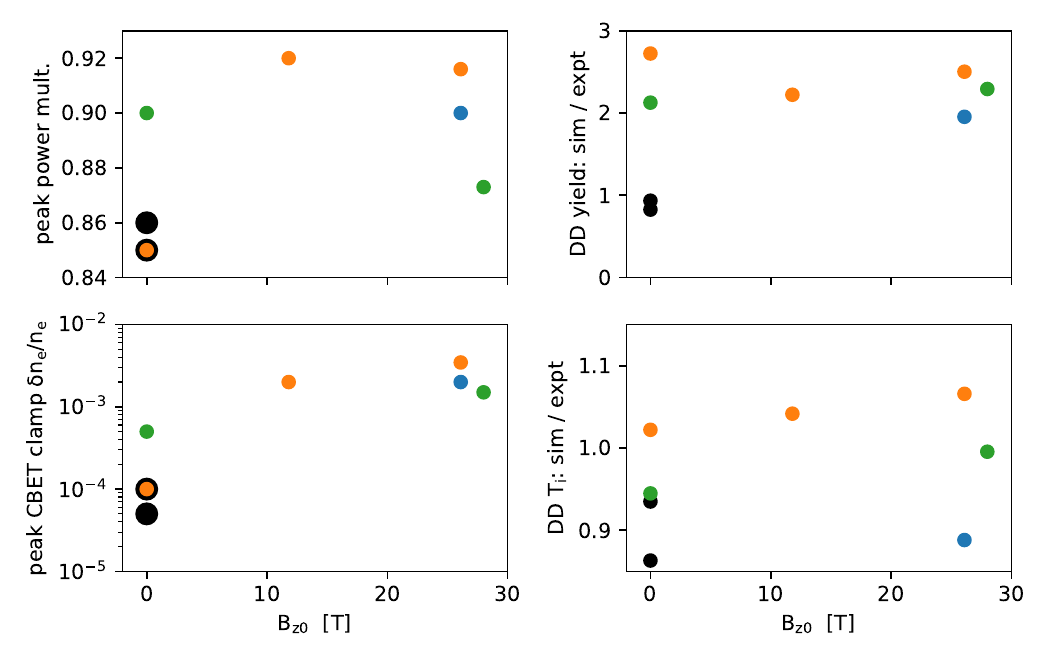}  
  \caption{Simulation tuned parameters (left) and comparison to experimental results (right) for NIF shots from Fig.\ \ref{fig:nifshots-expt}.}
  \label{fig:nifshots-sim}
\end{figure*}

N190721-2 unfortunately used 1.3 MJ of laser energy and is beyond our SBS risk limit.  We base our initial design on an earlier BigFoot series with slightly less laser energy, specifically gas-filled capsule shot N161204-3.  Table \ref{tab:shots} and Figs.\ \ref{fig:gascap-pulses}-\ref{fig:nifshots-sim} present details on the shots discussed here. The first two WarmMag shots, N201228-1 ($B_{z0}=0$) and N210301-1 ($B_{z0}=26$ T), stayed as close to N161204-3 as our constraints allowed (except for the reduced capsule fill density).  The differences of this initial low-T platform vs.\ N161204-3 are:
\begin{itemize}
  \item Reduced late-time laser power to meet SBS risk limit, see Fig.\ \ref{fig:gascap-pulses}.
  \item Changed capsule gas fill from 6.74 mg/cm$^3$ of D$_3$-$^3$He$_7$ to 5.10 mg/cm$^3$ of D$_3$-$^4$He$_7$ to give more convergence and B-field compression.
    \item Changed hohlraum gas fill from 0.3 mg/cm$^3$ of He to 0.258 mg/cm$^3$ of C$_5$H$_{12}$ to prevent LEH window bursting.
    \item Changed hohlraum wall from 30-50 $\mu$m thick Au to 15 $\mu$m thick AuTa$_4$ to reduce electrical conductivity and field soak-thru issues.
    \end{itemize}

The goals of the WarmMag NIF shots were to demonstrate: 1.\ that hohlraum x-ray conversion was similar with or without a B field, and with an AuTa$_4$ instead of pure Au hohlraum; 2.\ a repeatable comparison of nuclear performance in shots with and without B, with a roughly round hot spot.  Our starting point was to hew as closely as possible to BigFoot shot N161204-3 given our constraints, and tuning implosion shape came later.  The first shot N201228-1 had no imposed field, and no capsule fill due to a fielding issue.  It was essentially a hohlraum coupling test.  The AuTa$_4$ hohlraum and modified laser pulse gave x-ray drive as measured by DANTE comparable to N161204-3 until the time of the caboose \cite{sio-maghohl-pop-2023}.  The backscatter was very low. The next shot N210301-1 had imposed field $B_{z0}=26$ T  and capsule fill. It gave similar x-ray drive and backscatter to the prior shot, and produced a very prolate implosion: $P_2/P_0=63\%$.  

\subsection{``WarmMag'' final low-T platform: round hotspot}

Our next goal was to make the magnetized implosion shape round and finalize the low-T platform.  We did this on the third shot N210607-2 by reducing the laser cone fraction, or ratio of inner-beam to total laser power.  We only reduced the inner beam power: the outer beams were already at their SBS limit, so we could not raise them to maintain total power.  We also changed the capsule fill to 3.99 mg/cm$^3$ of pure D, to increase yield via increased D density and reduced hotspot radiative loss. This design was modeled in Lasnex to give a round implosion, which was basically achieved: $P_2/P_0=5.7\%$.  Our next goal was an unmagnetized shot of this design.  Several attempts had fielding issues, with the best one being N210912-1 (which still dropped 2 of the 48 NIF laser quads).  We subsequently shot an intermediate $B_{z0}= 12$ T on N220912-1, which had almost the same yield and $T_i$ as $B_{z0}= 26$ T.  This validates the expected performance scaling for gas-filled targets where reduced electron thermal conduction is the main B-field effect \cite{walsh-magicf-pop-2022}.

\begin{table*}    
  \centering
  \begin{tabular}{|c|c|c|c|c|c|c|c|c|}
    \hline
    NIF Shot                         & N161204-3       & N161205-3   &  N210301-1      & N210912-1 & N220912-1  & N210607-2  & N230612-1  & N230212-2   \\
    \hline
    Shot parameters                  &&&&&&&&   \\
    Platform                         & BF              & BF          & WM first low-T  & WM low-T  & WM low-T   & WM low-T   & WM high-T  & WM high-T   \\
    $B_{z0}$ [T]                     & 0               & 0           & 26.1             & 0         & 11.8       & 26.1      & 0          & 28.0         \\
    Laser energy [kJ]                & 1092            & 1065        & 926             & 840        & 874        & 883       &  940       & 989          \\
    Cone fraction (inner/total)      & 27.3            & 26.7        & 27.2            & 21.2       & 22.1       & 21.9      & 25.7      & 27.0          \\
    LEH diameter [mm]                & 3.45            & 3.45        & 3.46            & 3.48       & 3.45       & 3.48      & 3.15       & 3.13         \\
    Capsule fill gas                 & D$_3$-$^3$He$_7$ & D$_1$-T$_1$ & D$_3$-$^4$He$_7$ & D         & D          & D         & D          & D            \\
    Capsule fill density [mg/cm$^3$] & 6.74            & 4.01         &    5.10         & 3.99      & 3.99      & 3.99       & 3.00      & 3.00          \\ 
    \hline
    Experimental data                &&&&&&&&  \\
    x-ray bangtime [ns]              & 7.1             & 7.04         & 7.48            & 8.00      & 7.58      & 7.65       & 7.53      & 7.40          \\
    x-ray $P_0$ [$\mu$m]             & 60              & 74.2         & 51              & 58.5      & 55.8      & 53.3       & 47        & 37.3          \\
    x-ray $P_2/P_0$ [\%]             & 18.7            & 8.1          & 63              & -16.6     & -17       & 5.7        & -24       & -5.8          \\
    DD neutron yield $Y$             & 8.69E11         & 2.60E12      & 5.00E11         & 6.72E12   & 1.71E13   & 1.95E13    & 1.03E13   & 2.26E13      \\
    DD $T_i$ [keV]                   & 3.07            & 3.38         & 4.2             & 2.72      & 3.6       & 3.8        & 3.44      & 4.4           \\
    \hline
    Lasnex modeling                  &&&&&&&&    \\
    Peak power multiplier            & 0.85            & 0.86         & 0.9             & 0.85      & 0.92      & 0.916      & 0.9       & 0.873         \\
    Peak CBET clamp $\delta n_e/n_e$ & 1E-4            & 5E-5         & 2E-3            & 1E-4      & 2E-3      & 3.46E-3    & 5E-4      & 1.5E-3        \\
    Inner $E_{post}/E_{inc}$ - 1 [\%] & 3.7             & 4.0          & 15.7           & 6.9        & 18.4      & 27.9       & 8.2       & 13.7          \\
    \hline
    Modeling vs.\ experiment        &&&&&&&&    \\
    bangtime: sim - expt [ps]       & 20              & 20           & 10              & 30        & 20        & -20        & 30        & -10          \\
    $P_2/P_0$: sim - expt [\%]      & 2.3             & 1.5          & 0               & 2.9       & -0.2      & 0.3        & 1.3       & 1.8          \\
    $Y$: sim / expt                 & 0.826           & 0.935        & 1.95            & 2.72      & 2.22      & 2.50       & 2.13      & 2.29         \\
    DD $T_i$: sim / expt            & 0.863           & 0.935        & 0.888           & 1.022    & 1.042      & 1.066      & 0.945     & 0.995        \\
    \hline
    Effect of B field               &&&&&&&& \\
    $Y[B]/Y[B=0]$: expt             & n/a             & n/a          & n/a             & 1: $[B=0]$  & 2.54   & 2.90        & 1: $[B=0]$  & 2.19        \\
    $Y[B]/Y[B=0]$: sim              & n/a             & n/a          & n/a             & 1           & 2.08   & 2.67        & 1           & 2.37        \\
    $T_i[B]/T_i[B=0]$: expt         & n/a             & n/a          & n/a             & 1           & 1.32   & 1.40        & 1           & 1.28        \\
    $T_i[B]/T_i[B=0]$: sim          & n/a             & n/a          & n/a             & 1           & 1.35   & 1.46        & 1           & 1.35        \\
  \hline
  \end{tabular}
  \caption{Experimental and Lasnex modeling data for NIF gas-filled capsule shots discussed in this paper.  For each platform, shots are ordered by $B_{z0}$ not shot date.  BigFoot (BF) shots all had Au hohlraum walls and hohlraum fills of 0.3 mg/cm$^3$ of He, while all WarmMag (WM) shots had AuTa$_4$ walls and fills of 0.258 mg/cm$^3$ of C$_5$H$_{12}$. ``Cone fraction'' is the inner-beam energy divided by total laser energy.  ``Inner $E_{post}/E_{inc}$'' is the post-CBET inner-beam energy divided by the incident and measures CBET from outer to inner beams.  Typical measurement error bars are $\pm50$ ps for bangtime, $\pm3\ \mu$m for $P_0$ and $P_2$, $\pm5\%$ for DD yield, and $\pm0.15$ keV for $T_i$.}
  \label{tab:shots}   
\end{table*}

\subsection{``WarmMag'' high-T platform: increased hotspot temperature}

In 2023 our goal was to increase the capsule hotspot temperature while still respecting our constraints.  This WarmMag ``high-T'' platform entailed four changes vs.\ the low-T one:
\begin{itemize}
  \item Extended caboose to increase laser energy and x-ray drive, see Fig.\ \ref{fig:gascap-pulses}.
  \item Reduced LEH diameter from 3.45 to 3.13 mm, to reduce hohlraum x-ray loss.
  \item Reduced capsule fill density from 3.99 to 3.00 mg/cm$^3$ to increase capsule convergence for a given x-ray drive.
  \item Increased laser cone fraction to maintain a round implosion shape.
\end{itemize}
These changes increased the hotspot temperature both with and without a B field, as demonstrated in our most recent shots N230212-2 and N230612-1.

Figures \ref{fig:nifshots-expt}-\ref{fig:nifshots-sim} provide a summary of the WarmMag NIF shots to date, with $B_{z0}$ the abscissa in all plots.  The delivered laser energy and cone fraction slightly increase with $B_{z0}$, which accounts for some of the decrease in bangtime with $B_{z0}$ (besides any field effects).  It also somewhat complicates drawing clear conclusions from the increase in $P_2$, yield and $T_i$ with $B_{z0}$.  A compelling demonstration of the benefits of magnetization is the higher yield and $T_i$ in the low-T, $B_{z0}=12$ T shot vs.\ the unmagnetized high-T shot: the field over-compensates for the higher laser energy, x-ray drive and capsule convergence.

\section{Lasnex Rad-MHD Modeling of WarmMag NIF Experiments}

We now describe our Lasnex rad-MHD modeling of the WarmMag NIF shots.  Hohlraum modeling at LLNL has traditionally not included MHD effects, though work is underway to address this deficiency.  Biermann-battery fields are known to be present in the hohlraum and capsule, and experiments are ongoing to understand their role.  Earlier hohlraum MHD modeling with the Hydra rad-MHD code \cite{marinak-hydra-pop-2001} includes studies of self-generated fields \cite{farmer-mhdhohl-pop-2017} and imposed fields \cite{strozzi-Bfield-jpp-2015}.  Compared to the earlier imposed-field work, the present work uses a more complete MHD model with current best practices, and considers targets with low hohlraum gas fill, HDC ablator and short laser pulses (the prior work considered a ``LowFoot'' design with plastic ablator, high hohlraum gas fill, and long laser pulses).  Appendix B presents details on MHD modeling with Lasnex and the LHT.  Imposed field is done through an initial condition on the vector potential $A_\phi[r,z]$, which produces $\vecB_{rz}$ but no $B_\phi$.  We use the analytic solution for a solenoid of finite axial length and no radial thickness \cite{caciagli-solenoid-jmagmat-2018}.  This idealization compares well to detailed calculations of the real NIF coils.

\begin{figure}
  \centering
  \includegraphics[width=3in]{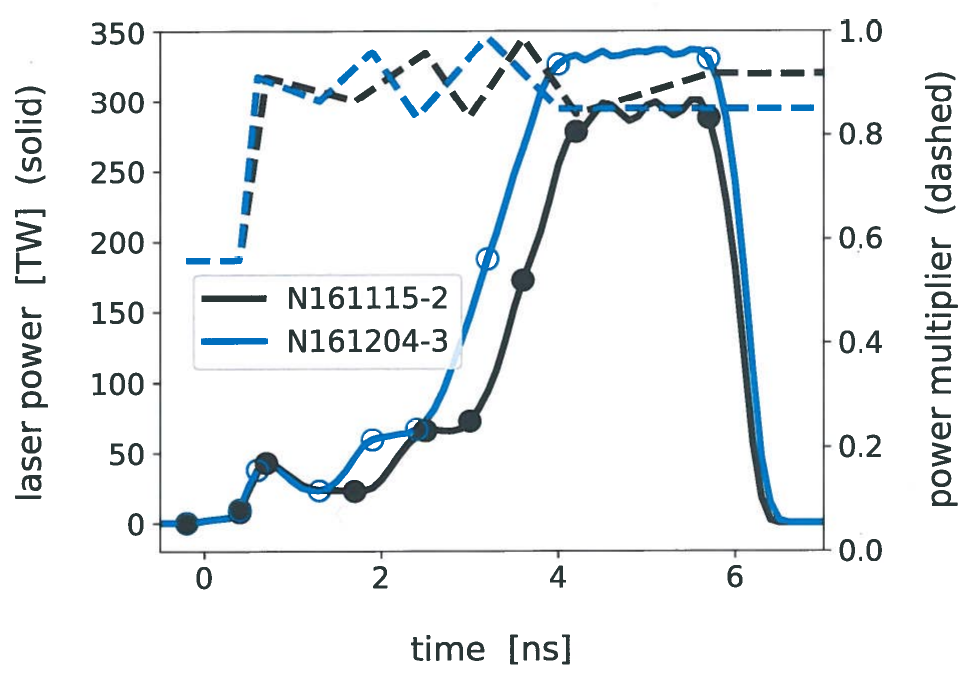}  
  \caption{Solid: incident laser power for BigFoot keyhole shot N161115-2 (black) and symcap shot N161204-3 (blue).  The circles indicate times for which ANTS finds power multipliers.  Dashed: ANTS power multipliers.}
  \label{fig:bf2016-keyhole-power}
\end{figure}

\begin{figure}
  \centering
  \includegraphics[width=3in]{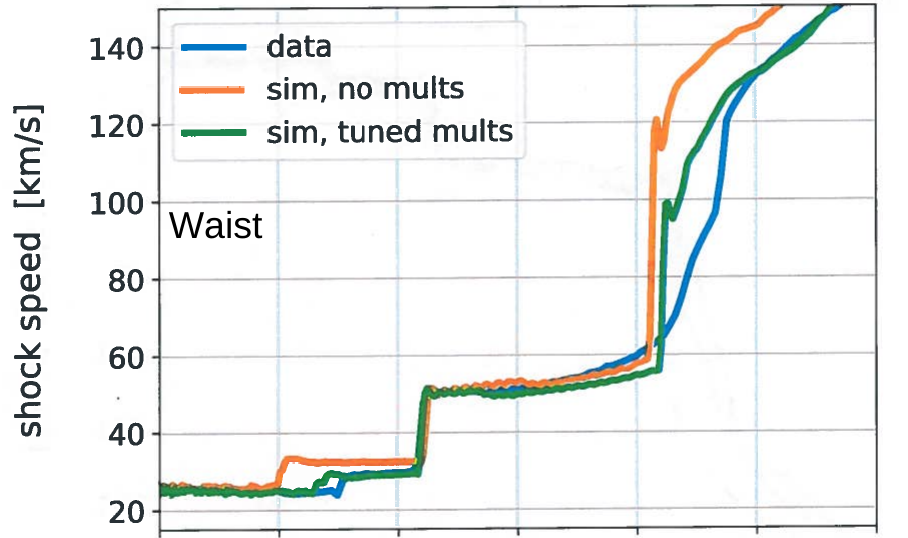}  
  \includegraphics[width=3in]{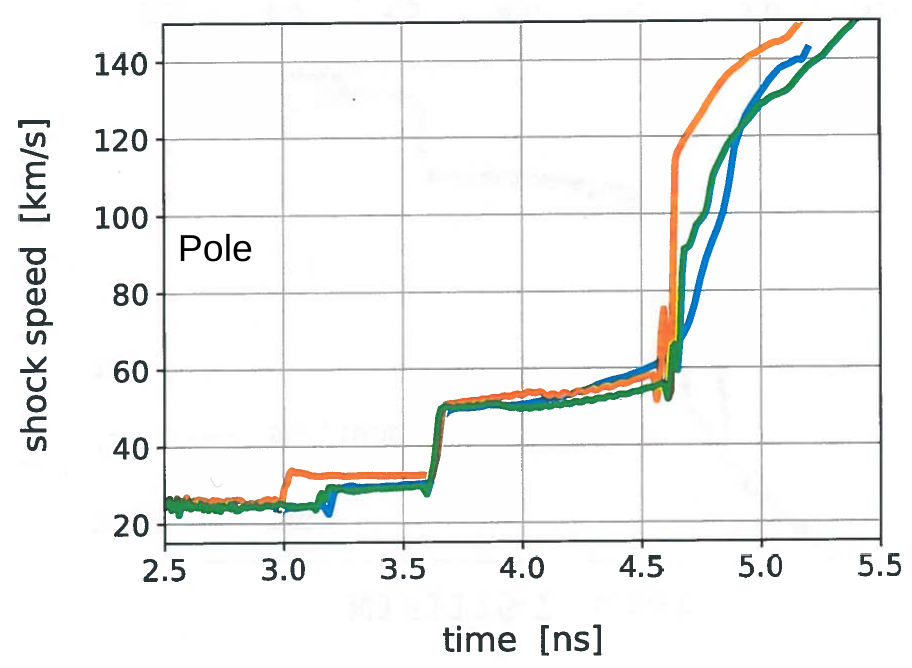}  
  \caption{Shock speed for BigFoot keyhole shot N161115-2, as measured along the capsule waist (top) and pole (bottom).  Blue is measured data.  Orange is LHT simulation with no power multipliers (i.e.\ a constant value of 1).  Green is LHT simulation with tuned power multipliers given in Fig.\ \ref{fig:bf2016-keyhole-power}.}
  \label{fig:bf2016-keyhole-shocks}
\end{figure}

Our modeling goals are to design the WarmMag NIF shots, and to understand how well we can model MHD effects.  We therefore follow the standard practice outlined in Appendix A of tuning laser power multipliers and the CBET saturation clamp to match data, namely shock velocity, bangtime and hotspot $P_2$ shape.  This provides what we consider to be a reasonably realistic x-ray drive for the capsule.  Subsequent capsule performance, such as yield and ion temperature, are untuned and thus an indication of simulation quality.

Shock velocity data is obtained from keyhole experiments, which can only be done cryogenically.  We therefore use the BigFoot keyhole shot N161115-2 to develop laser power multipliers.  This shot directly applies to the two BigFoot gas-filled shots we model, but not to the WarmMag platform (with different e.g.\ wall material and hohlraum fill).  We use the Automated NIF Tuning Suite (ANTS) \cite{weber-ants-popprep-2024} to develop our tuned parameters.  Figure \ref{fig:bf2016-keyhole-power} plots the delivered laser power and multipliers, and Fig.\ \ref{fig:bf2016-keyhole-shocks} shows the resulting shock velocities. The power multipliers clearly improve the agreement between simulated shocks and data, but it is not perfect.  We deem this to be adequate for the present study of gas-filled capsules, which should be less sensitive to shock dynamics than DT ice-layered implosions. The power multipliers for N161115-2 are applied to all other shots, at equivalent times found by ANTS.  Figure \ref{fig:bf2016-keyhole-power} illustrates this for N161204-3.  A single peak power multiplier is chosen for each shot other than N161115-2 to match the implosion bangtime.  For N161204-3 this starts around 4 ns.  The CBET clamp $\delta n_e/n_e$ is 0.005 until 3.8 ns (near the end of the rise to peak power), which is a typical value used for similar shots.  After 3.8 ns a new clamp is chosen to match the hotspot $P_2$.  For all shots, the simulated bangtime is within 30 ps of data, and $P_2/P_0$ is within 2.9\%.

\begin{table*}    
  \centering
  \begin{tabular}{|c|c|c|c|c|c|c|}
    \hline
    Run: MHD model            & 1: No MHD & 2: self-gen, $f_N=1$ & 3: self-gen, $f_N=0.1$ & 4: 3 plus $B_{z0}=30$ T & 5: 3 capsule-only & 6: 4 capsule-only \\
    \hline
    Peak $T_{rad}$ [eV]       & 287.91 & 288.81                & 288.97                & 288.20                & n/a               & n/a        \\
    Peak M-band fraction [\%] & 18.60  & 18.62                 & 18.68                 & 18.44                 & n/a               & n/a        \\
    x-ray bangtime [ns]       & 7.25   & 7.20                  &  7.19                 & 7.21                  & 7.16              & 7.16      \\
    x-ray $P_0$ [$\mu$m]      & 58.15  & 58.68                 & 58.39                 & 56.35                 & 58.54             & 57.88     \\
    x-ray $P_2/P_0$ [\%]      & 2.06    & 12.90                & 19.97                 & 16.17                  & -0.20             & 9.73       \\
    DD neutron yield          & 9.94E11 & 7.34E11             & 7.20E11                & 1.12E12                & 7.42E11           & 1.05E12    \\
    DD $T_i$ [keV]            & 2.96    & 2.63                 &  2.64                 &  3.53                  & 2.67             & 3.40       \\
    \hline
  \end{tabular}
  \caption{Lasnex modeling of BigFoot shot N161204-3 with different MHD models.  ``Self-gen'' means run only had self-generated $B_\phi$ but no imposed fields.  $f_N$ is the Nernst multiplier outside the capsule ($f_N=1$ inside the capsule).  $B_{z0}$ is the initial imposed field at capsule center.  ``Capsule-only'' runs 5 and 6 do not include the hohlraum and are driven by symmetrized versions of run 3's x-ray drive.}
  \label{tab:n161204}   
\end{table*}

\begin{figure*}
  \centering
  \includegraphics[width=7.25in]{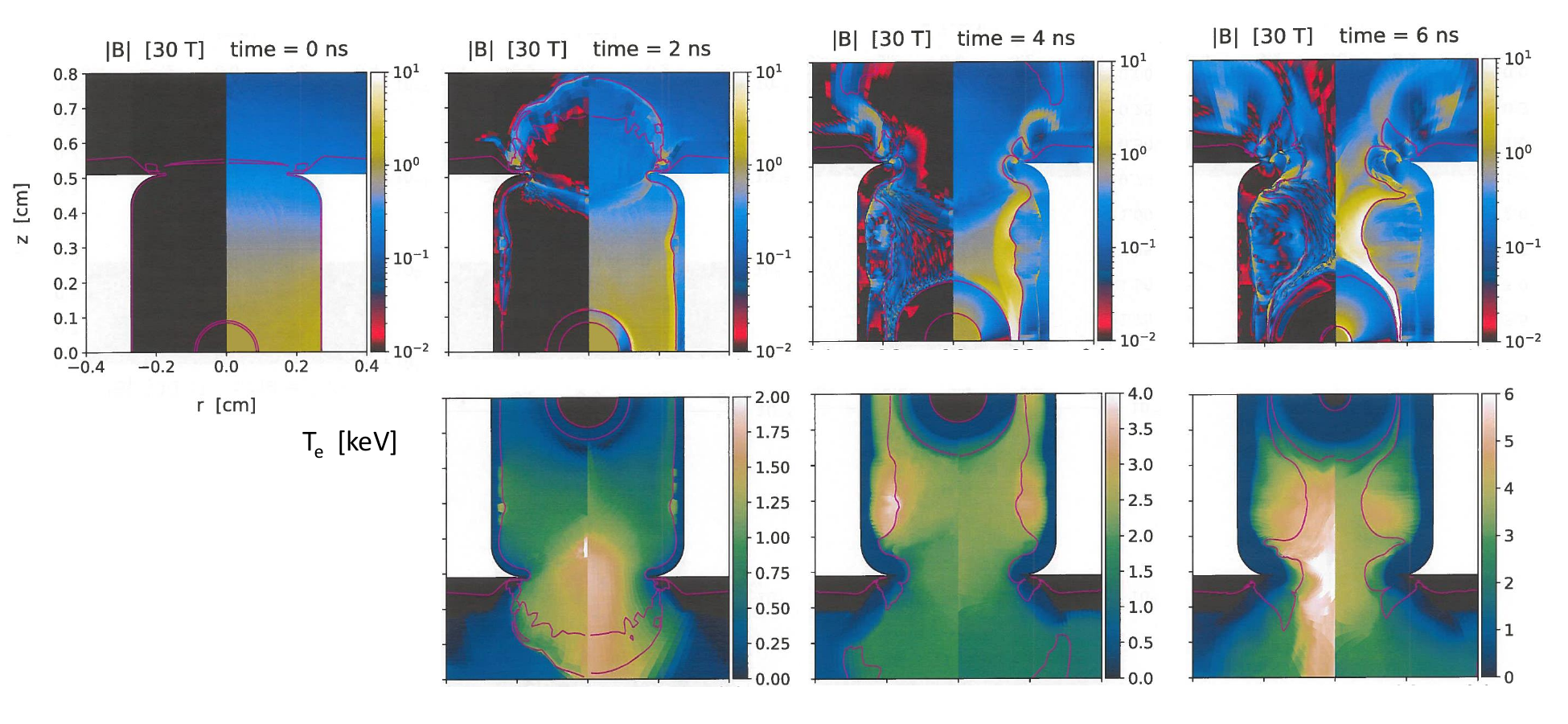}
  \caption{BigFoot shot N161204-3 Lasnex modeling from Table \ref{tab:n161204}: hohlraum $|\vecB|$ (top) and $T_e$ (bottom) at several times.  Left half $r<0$ is for run 3 with self-generated field and $f_N=0.1$, and right half $r>0$ is for run 4 with initial $B_{z0}=30$ T field.  $|\vecB|$ is scaled to 30 T.  The imposed-field run is hotter in the LEH at 2 ns and 4 ns, but at 6 ns it is hotter in the self-generated run.}
  \label{fig:bigfoot-hohl2d}
\end{figure*}

We return to Fig.\ \ref{fig:nifshots-sim} to discuss the modeling results.  The peak power multipliers vary from 0.85 to 0.92, without a clear trend with $B_{z0}$: the multiplier increases toward unity for the low-T shots but decreases for the high-T ones.  All WarmMag shots have multipliers equal to, or closer to unity than, BigFoot.  The peak CBET clamp increases substantially with $B_{z0}$, from 5E-5 -- 5E-4 for unmagnetized shots to 1.5E-3 -- 3.5E-3 for magnetized ones (recall a larger clamp means \textit{less} artifical modification to the physical CBET model).  This results in significantly more CBET to the inner beams with increasing $B_{z0}$: see Table \ref{tab:n161204}.  Physical ion-wave nonlinearities can limit CBET for amplitudes $\delta n_e/n_e \gtrsim 0.01$.  All of this indicates the WarmMag platform and B fields do not make hohlraum modeling less accurate (and thus requiring more tuning to match data).  With tuning parameters chosen, the capsule yield and $T_i$ are untuned and reflect simulation quality.  The $T_i$ simulated / experiment ratio is closer to unity (at most 1.066) for all but one WarmMag shot than the average of the two BigFoot shots (0.904).  The exception is N210301-1 (the blue point: 0.888) which was extremely prolate and could have $T_i$ compromised by substantial plasma flow.  The relative yield increase due to magnetization (see Table \ref{tab:shots}) is $2.19-2.90\times$ in the data, and is similar but slightly lower on average in the modeling $(2.08-2.67\times)$.  For $T_i$ the relative increase due to magnetization is $1.28-1.4\times$ in the data and slightly higher $(1.35-1.46\times)$ in modeling.  The yield and $T_i$ increases are roughly in line with modeling, especially given measurement error bars.

The major shortcoming of the modeling is that the absolute simulated yield is $1.95-2.72\times$ the data in the WarmMag platform, while it agrees well for BigFoot $(0.83-0.94\times$).  The discrepancy is slightly less for the magnetized shots, which indicates the field is at least not introducing new mysteries.  This suggests the field may reduce capsule mix, but this is speculative.  There are many differences between the BigFoot and WarmMag platform which could account for the difference in yield modeling.  We are exploring these now and will report results in the future.  In particular, modeling of capsule mix due to hydro instabilities, the fill tube, etc.\ remains to be done.

\section{Rad-MHD Study of NIF BigFoot Experiment N161204-3}

This section studies the role of MHD and imposed B fields in BigFoot shot N161204-3.  This is closer to a typical cryogenic NIF implosion than our WarmMag shots: it uses a pure Au hohlraum, He hohlraum fill, and no caboose on its laser pulse. We use our Lasnex modeling methodology and only vary the MHD physics included.  We consider four integrated hohlraum modeling runs: 1: a base run with no MHD; 2: run 1 with self-generated azimuthal field and the full Nernst effect (Nernst multiplier $f_N=1$, see Eq.\ \ref{eq:Rbeta}); 3: run 2 but with our standard $f_N=0.1$ (except $f_N=1$ inside the capsule); and 4: run 3 but with a finite-solenoid imposed $\vecB_{rz}$ field and $B_{z0}=30$ T.  All runs use the same laser power multipliers and CBET clamp as given in Table \ref{tab:shots} for this shot, which corresponds to run 3.

Figure \ref{fig:bigfoot-hohl2d} shows $|\vecB|$ and $T_e$ in the hohlraum at several times.  Self-generated field is produced mostly by the Biermann battery effect (misaligned $\nabla n_e$ and $\nabla T_e$).  This leads to highly localized field, such as at the LEH window boundary (2 ns) and where the laser is absorbed near the hohlraum wall.  The imposed $B_{z0}=30$ T field roughly follows the frozen-in law: the field increases where plasma compresses normal to $\vecB$ (like the hohlraum fill) and decreases where it expands (like the ablated hohlraum wall and capsule).  The imposed $\vecB_{rz}$ field roughly ``adds on'' to the self-generated field.  $T_e$ is hotter for the imposed-field run  in the LEH at 2 ns, as expected due to reduced electron thermal conduction.  This difference is reduced at 4 ns, and the ``gold bubble'' where the outer beams hit the wall is hotter in the self-generated run.  At 6 ns the hohlraum fill is generally hotter in the self-generated run.  This goes against the expectation that the main effect of the imposed field is to reduce thermal conduction.  It also seems to contradict $T_e$ measurements in hohlraums at the OMEGA Laser Facility \cite{montgomery-magnetized-pop-2015}.  Those hohlraums were much smaller in scale than our NIF experiments: they were driven by $\sim1/40$ the laser energy, had equivalently shorter spatial dimensions and laser pulses, and had no capsule.  In absolute terms, those experiments resemble the LEH early in time of the NIF shots, which are hotter with the imposed field.  Analysis of the later times when the imposed-field run is cooler is ongoing.  One possibility is the higher fill pressure, from both higher $T_e$ and magnetic pressure, reduces compression and $pdV$ heating.

\begin{figure}
  \centering
  \includegraphics[width=3in]{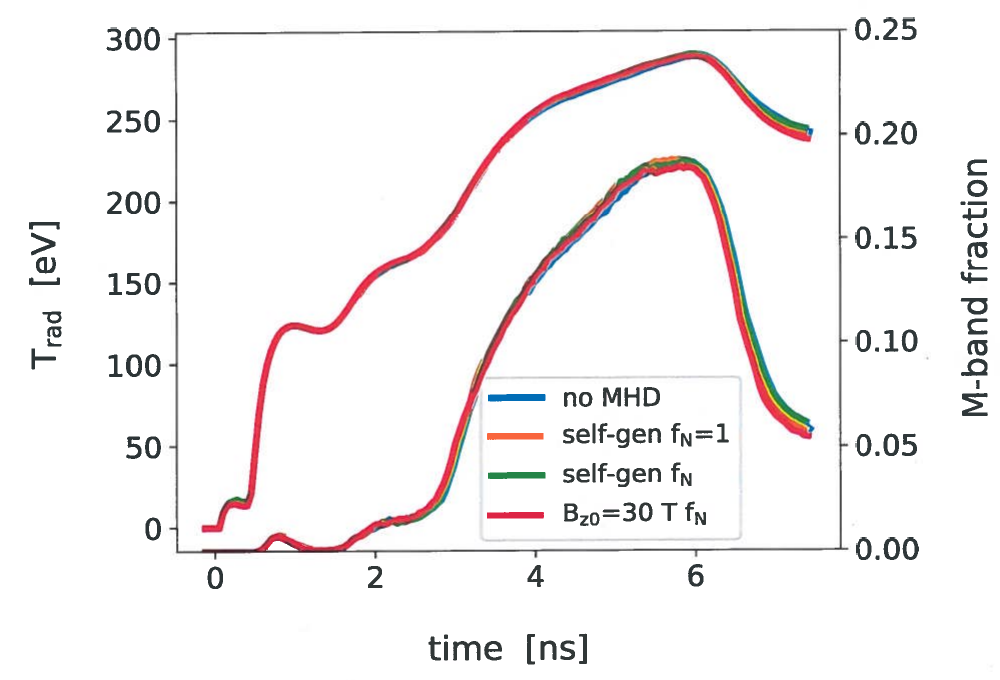}
  \caption{BigFoot shot N161204-3 Lasnex modeling: x-ray drive along a contour enclosing the capsule.  $T_{rad}$ is the spectrally-integrated radiation temperature, and M-band fraction is the spectral fraction with $h\nu \geq 1.8$ keV.  Blue: run 1 with no MHD.  Orange: run 2 with self-generated B fields and no Nernst multiplier ($f_N=1$).  Green: run 3: run 2 but standard LHT $f_N=0.1$ outside capsule.  Red: run 4: run 3 plus initial solenoidal $B_r/B_z$ field with $B_{z0}=30$ T at capsule center.  There is very little difference between the four runs.}
  \label{fig:bigfoot-Trad-Mband}
\end{figure}

\begin{figure}
  \centering
  \includegraphics[width=3in]{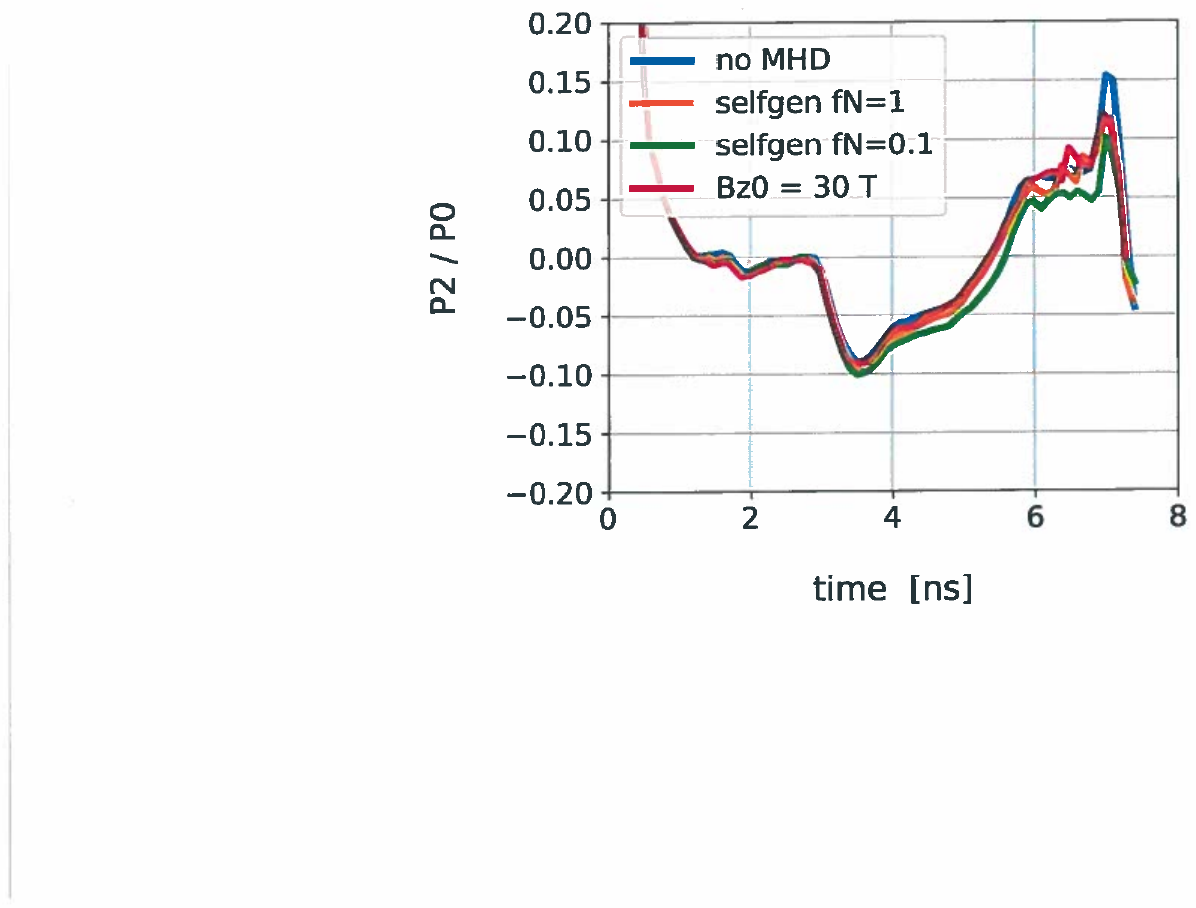}
  \caption{BigFoot shot N161204-3 Lasnex modeling: $P_2$ Legendre moment of the hohlraum x-ray flux at a sphere of radius 100 $\mu$m larger than the time-dependent capsule ablation front.  Same simulations and coloring as Fig.\ \ref{fig:bigfoot-Trad-Mband}.}
  \label{fig:bigfoot-rmfmom-p2}
\end{figure}

The hohlraum x-ray drive as seen by the capsule is presented in Fig.\ \ref{fig:bigfoot-Trad-Mband}.  There is very little difference in total x-ray flux or the M-band fraction (photon energy $>1.8$ keV) between the four runs.  MHD and imposed field thus have little effect on overall hohlraum energetics. Fig.\ \ref{fig:bigfoot-rmfmom-p2} shows the $P_2$ Legendre moment of the x-ray flux.  We follow the standard sign convention with $P_2>0$ for a pole-hot x-ray drive, which produces an oblate ($P_2<0$) hotspot shape.  The four runs are similar but not identical: the no-MHD run has the most pole-hot drive, the self-generated with $f_N=0.1$ the least, and the other two runs are in between. The peak-power CBET is very small in all four runs since we used the very low clamp of 1E-4 from our tune for this shot.  A better CBET model without the need for a clamp may reveal differences between the runs.

\begin{figure}
  \centering
  \includegraphics[width=3in]{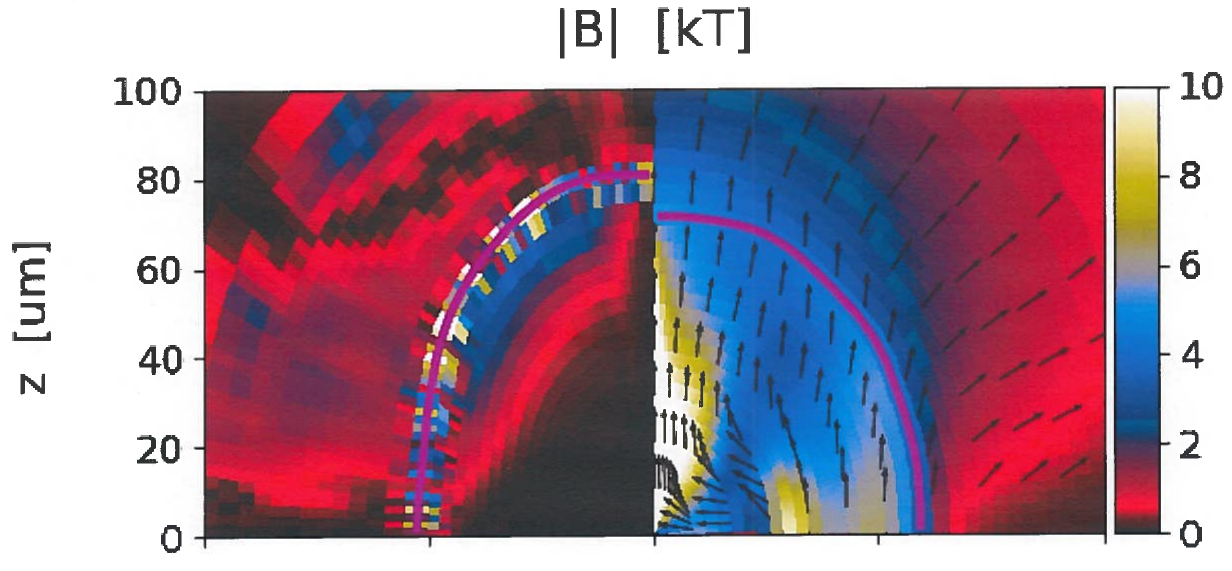} \\
  \includegraphics[width=3in]{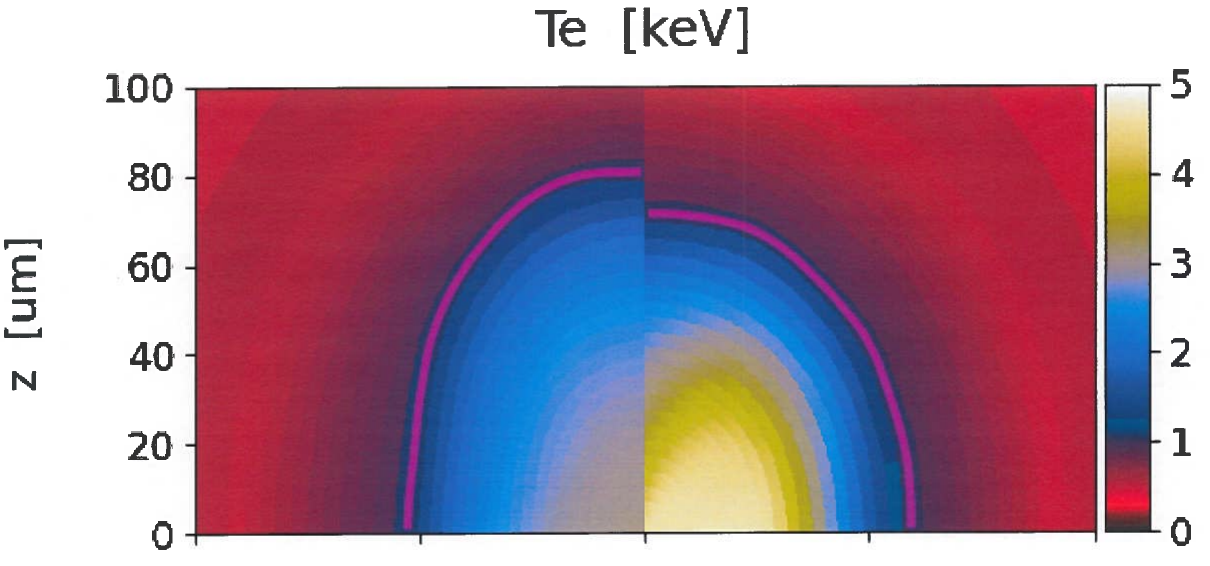} \\
  \includegraphics[width=3in]{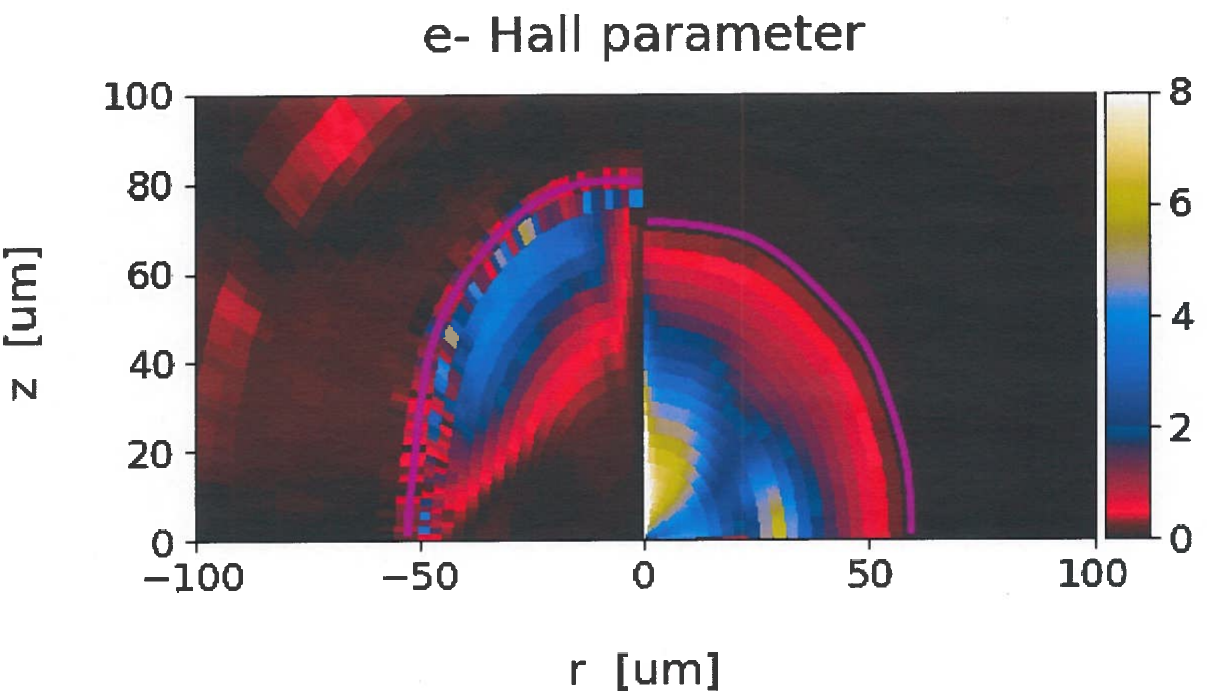}
  \caption{BigFoot shot N161204-3 Lasnex modeling: simulated conditions at 7.25 ns (slightly after bangtime) for run 3 (left: no imposed field) and 4 (right: imposed field).  Magenta contour is the capsule gas - ablator interface. Top: $|\vecB|$ including all three field components; left panel: $|B_\phi| \times 100$ $(B_r=B_z=0)$, right panel arrows indicate direction of $(B_r, B_z)$.  Middle: electron temperature $T_e$.  Bottom: electron Hall parameter $\om_{ce}\tau_{ei}$, left panel $\times1000$.}
  \label{fig:bigfoot-cap2d}
\end{figure}

We now discuss the capsule physics of the four hohlraum runs, as seen in Table \ref{tab:n161204} and Fig.\ \ref{fig:bigfoot-cap2d}.  The most salient effect of self-generated fields, as seen in comparing runs 2 and 3 vs.\ 1, is the less pole-hot x-ray flux $P_2$. This is consistent with the hotspot $P_2$ for runs with self-generated fields being prolate vs.\ the round no-MHD run.  The very low capsule Hall parameter < 0.01 in Fig.\ \ref{fig:bigfoot-cap2d} suggests that self-generated capsule fields have very little effect. This is as expected: high-mode asymmetries are required to generate significant magnetic flux in capsules \cite{walsh-biermann-pop-2021}, and are not included here. The bangtimes are all close (within 60 ps), consistent with the close x-ray fluxes. The lower yield and $T_i$ vs.\ no-MHD is likely due to the prolate shape.  The only difference between the two self-generated runs with different Nernst multiplier is in hotspot $P_2$, again consistent with the x-ray flux $P_2$.

The main effect of the imposed $B_{z0}=30$ T field in run 4 is to increase yield by 56\% and $T_i$ by 34\% vs.\ run 3.  The x-ray flux $P_2$ is slightly more pole hot, which is consistent with the slightly lower hotspot $P_2$.  This is in contrast to a na\"ive expectation that an axial field leads to a more prolate hotspot, since electron thermal conduction is unmagnetized along $z$.  To separate the effects of the imposed field on hohlraum x-ray drive and capsule physics, we performed runs 5 and 6.  These are capsule-only runs both driven by the x-ray flux from hohlraum run 3.  This is done via a frequency-dependent source (FDS) imposed outside the capsule vs.\ time and photon energy, but symmetric in space.  Any asymmetry is the result of capsule physics.  Run 5 with no imposed field has $P_2/P_0=-0.2\%$, which is very small and indicates residual numerical asymmetry in either the simulation or post-processing.  The imposed-field run is prolate ($P_2/P_0=9.7\%$), as expected.  In the imposed-field hohlraum run, the more pole-hot x-ray flux over-compensates for this effect.

\section{Conclusions and Future Work}

This paper reviewed the magnetized NIF hohlraum experiments to date with room-temperature, gas-filled capsules, which we call the WarmMag platform.  The emphasis was on the design considerations and modeling with the Lasnex rad-MHD code and the LHT common model.  The WarmMag platform is based on BigFoot gas-filled capsules driven by $1.05-1.1$ MJ of laser energy.  Changes were made to satisfy the constraints of allowing the field to soak thru the hohlraum wall without excessive $J\times B$ motion (novel AuTa$_4$ hohlraum instead of standard Au or DU), room-temperature fielding (hohlraum fill gas of C$_5$H$_{12}$ instead of He), and a stringent laser energy limit $\lesssim 1$ MJ to avoid SBS optics damage risk.  The two most recent shots in the ``high-T'' campaign on the WarmMag platform, N230212-2 and N230612-1, were presented for the first time.  These increased the capsule compression and hotspot temperature with or without an imposed field, and showed an imposed field gives a hotter hotspot.  A compelling demonstration of the benefits of magnetization is that the lowest imposed field of 11.8 T in the ``low-T'' shot N220912-1 gave a higher neutron yield and ion temperature $T_i$ than the best-performing unmagnetized shot N230612-1 in the high-T campaign.

Lasnex modeling captures most of the NIF observations, with the notable exception of the absolute neutron yield being $2.0-2.7\times$ higher than the data.  Hohlraum modeling required similar laser power multipliers to match the implosion bangtime for WarmMag shots with and without imposed field.  The needed clamp $\delta n_e/n_e$ on cross-beam energy transfer (CBET) to match hotspot $P_2$ increased greatly with $B_{z0}$, from 5E-5 -- 5E-4 for unmagnetized shots to 1.5E-3 -- 3.5E-3 for magnetized ones.  This is approaching the level of 0.01 when physical ion-wave nonlinearities may play a role.  With hohlraum tuning parameters (power multiplier and CBET clamp) chosen, the capsule outputs of yield and $T_i$ are untuned and indicate simulation quality.  The modeling is close to the measured $T_i$ and relative yield and $T_i$ increase due to magnetization, but the modeled yields are too high.  We will study yield degradation due to mix, hydro instabilities and the fill tube in the future, all of which our current modeling neglects.

A modeling study of effects from self-generated (mostly Biermann battery) and imposed fields ($B_{z0}=30$ T) on BigFoot shot N161204-3 showed both cause little change in the total x-ray drive and its M-band fraction.  There is some effect on the x-ray drive's $P_2$ moment, with the imposed field making it slightly more pole-hot.  This leads to a slightly less prolate hotspot vs.\ the self-generated run, though capsule-only modeling with the same, symmetric x-ray drive shows the imposed field makes the hotspot more prolate.  The $P_2$ variations due to imposed field in both the x-ray drive and hotspot shape are minor, and should be tunable through standard techniques like laser cone fraction and wavelength shift (to control CBET).  The optimal shape of a magnetized implosion is an open question, but it is likely not too far from round.  Non-spherical implosion comes at a substantial cost in reduced compression: the limit of a cylindrical implosion gives a density increase $\sim CR^2$ not $\sim CR^3$ for a spherical one (CR is convergence ratio).

To close we discuss ongoing work at LLNL towards magnetized indirect-drive ignition on NIF.  Current experiments are limited by being room-temperature and having $\lesssim 1$ MJ of laser energy, so magnetized ignition work is mostly theory and modeling.  This is broadly geared toward finding the best uses of imposed field for ignition designs.  The main thrust is to increase the maximum achievable yield and gain on NIF.  We start by imposing fields with $B_{z0} \leq 100$ T on the best performing NIF implosions, in the Hybrid E campaign.  Early modeling results show up to tripling the yield with $B_{z0} \sim 60-70$ T.  We are also investigating magnetized target designs for use on NIF ``Enhanced Yield Capability'' (EYC) which is a proposal for upgrading NIF to $2.6-3$ MJ of laser energy.  As mentioned above, a cryogenic pulsed-power system which mitigates the SBS optics risk due to prefire has been designed, but its construction timeline is uncertain.  Ongoing research on advanced hohlraum coils and pulser architecture are showing promise in achieving higher magnetic fields and easier target fabrication.

Future work will consider variants on these designs optimized to take advantage of the field, and other designs entirely.  For instance, we speculate that a field may give the most benefit for designs that do not have a hot enough hotspot to self-heat and ignite without a field.  We have also studied non-axial fields, such as magnetic mirrors or even closed field lines, which can give better performance.  A major challenge is how to impose such field of reasonable magnitude without seriously disrupting the implosion (say by running a wire through the capsule).  Besides increasing the maximum yield, another goal is to achieve the same yield but with relaxed requirements on laser energy or target quality.  This is appealing for ``users'' of ignition, who desire a $1-10$ MJ yield to conduct other high-energy-density (HED) experiments.

An important topic we are starting to address is the role of magnetic field in capsule degradation (``mix'') due to hydro instabilities or engineering features like the fill tube or capsule support tent.  Some experimental \cite{samulski-magrt-mre-2022} and simulation \cite{manuel-magmix-mre-2021} work exists, but this topic is little studied in ignition-relevant implosions.  We are conducting more gas-filled capsule implosions on the NIF WarmMag platform to study how imposed field alters mix.  These use the dual crystal backlighter imager (CBI) platform \cite{hall-cbi-pop-2024} to image high-Z tungsten x-ray emission from the implosion hotspot.  Experiments at the OMEGA laser are underway to study magnetized mix, as well as magnetized CBET.    Recent 3D MHD modeling of the Rayleigh-Taylor instability \cite{walsh-magRT-pre-2022} in idealized settings shows reduced growth for k-vectors $\veck || \vecB$ but not for $\veck \perp \vecB$.  In this case, growth can be enhanced by reduced thermal conductivity, which reduces ablative stabilization \cite{walsh-magmix-pop-2019}.  We plan to extend this work to integrated implosion modeling.  A very desirable goal is to develop reduced mix models like those available in Lasnex (discussed in Appendix A and Refs.\ \onlinecite{hurricane-richtmyer-pof-2000, zhou-buoyancy-drag-pre-2002, bachmann-mix-pop-2023}) to include magnetization.

We look forward to this and related work culminating in magnetized ignition experiments on the NIF.

\begin{acknowledgments}

It is a pleasure to thank C.\ A.\ Thomas and K.\ L.\ Baker for fruitful discussions about the BigFoot platform.  We are grateful to the NIF and OMEGA operations teams for fielding our nonstandard experiments, and to the target fabrication team for providing AuTa$_4$ hohlraums.  This work was performed under the auspices of the U.S.\ Department of Energy by Lawrence Livermore National Laboratory under Contract No.\ DE-AC52-07NA27344.  Partially supported by LLNL LDRD projects 20-SI-002 and 23-ERD-025.

\end{acknowledgments}

\section*{Data Availability}

The data that support the findings of this study are available from the corresponding author upon reasonable request.

\appendix

\section{Lasnex Hohlraum Template (LHT) Rad-MHD Simulation Model}

The rad-MHD modeling presented here uses the Lasnex simulation code \cite{zimmerman-lasnex-cppcf-1975, harte-lasnex-icf-1996}, which is a 2D axisymmetric ($rz$), arbitrary-Lagrangian-Eulerian (ALE) code.  We use the Lasnex Hohlraum Template (LHT), a ``best-effort,'' version-controlled common model used by the LLNL ICF program.  An earlier version of this model is described in Ref.\ \onlinecite{jones-hohlraums-pop-2017}.  We describe the current recommended practice for ``design-quality'' modeling of NIF hohlraums.  A typical LHT 2D hohlraum run takes 12 to 48 wall-clock hours on 36 Intel CPUs.  The LHT also supports 1D and 2D capsule-only runs, driven either directly by a laser, an x-ray frequency-dependent source with no angular spatial variation, or an x-ray ``tallied flux'' source with such variation.  The external laser entrance hole (LEH) hardware is included, namely the retainer ring and washer that secure the main window and thin secondary ``storm window'' which prevents NIF chamber gas from condensing.  The LEH hardware can affect LEH plasma conditions and thus cross-beam energy transfer (CBET), which transfers energy among the NIF lasers \cite{higginson-nearvachohl-pop-2022}.  The storm window itself is not currently included, due to mesh management challenges.

\textbf{Laser model:} The LHT treats the lasers with 3D ray tracing with an effectively infinite speed of light: at each time step, rays propagate through the target until the remaining power is $<$ 1E-4 of incident.  This includes refraction and inverse bremsstrahlung absorption as reduced by the Langdon effect \cite{langdon-invbrem-prl-1980}.  The incident ray positions, k-vectors and relative powers are chosen to accurately represent the near-field square lens aperture, far-field intensity profile due to phase plates, and pointing of each of the 192 NIF beams (or 96 in the more common one-sided runs shown here).  The NIF beams are grouped into ``quads'' of 4 beams.  The LHT follows this in that rays for the 4 beams in a quad are treated as a single laser with one power history, and function as a unit for CBET.  The inline CBET model is used \cite{strozzi-inline-prl-2017}, which currently treats the lasers as unpolarized quads and finds their 3D intensity on an auxiliary 3D mesh (the plasma conditions are still axisymmetric).  The quad treatment involves $\sim16\times$ fewer couplings than the beam treatment and is thus much faster.  The inline CBET model currently does not handle polarization, which an accurate beam model needs.  CBET between two lasers transfers power to the one with the longer wavelength in the plasma-flow frame, which depends on both flow-induced Doppler shifts and differences in the incident (lab-frame) wavelengths.  All four NIF cones (located at polar angles of 23, 30, 44 and 50 degrees) can have separate wavelengths.  All shots discussed in this paper have one wavelength for all cones.  Inline stimulated Brillouin (SBS) and Raman (SRS) backscatter models are available \cite{strozzi-inline-prl-2017}, but not used in this paper: backscatter for the relevant shots is generally low.

\textbf{Atomic and plasma models:} Material zones are treated with local thermodynamic equilibrium (LTE) models for properties like x-ray opacity and equation of state, until they exceed a critical electron temperature $T_{cr}$ of 0.3 keV.  We raise this to 1 keV ``inside the capsule'', meaning all capsule regions other than the outermost one: the latter ablates into the hohlraum, is heated by the laser, and becomes part of the hohlraum fill. The higher $T_{cr}$ reduces computer cost and agrees with current LLNL practice for high-resolution capsule instability modeling. Above $T_{cr}$, all properties come from LLNL's detailed configuration accounting (DCA) non-LTE atomic models \cite{jones-hohlraums-pop-2017, scott-nlte-hedp-2010}.  DCA is not a single model but a framework which accommodates models of widely varying complexity and cost. The results shown here use the ``November 2020'' DCA models with detail considered acceptable for evolving atomic populations ``inline'' in current ICF rad-hydro simulations.  We also use a tabulated, steady-state non-LTE approach based on the Linear Response Method in the high-Z hohlraum wall \cite{scott-nltetables-pop-2022}.  These tables are generated with substantially more detailed models than the inline ones, and greatly reduce the non-LTE computational cost.

The multi-ion species hydrodynamics package is used \cite{higginson-nearvachohl-pop-2022}, run here in an intermediate mode where each ion species has its own density and flow velocity but a single ion temperature (to reduce computer time and numerical issues).  The package supports per ion species temperatures and heat fluxes.  The plasma anisotropic stress tensor is also used.  Electrons are not treated as an independent species, instead the plasma is assumed to be quasi-neutral: $n_e=\sum_i Z_in_i$.  Separate electron and ion temperatures are used, and the net current in the MHD package is equivalent to a separate electron flow velocity.  Ion-ion transport and coupling coefficients come from Ref.\ \onlinecite{stanton-ion-xport-pre-2016}, while electron-ion (e-i) coefficients come from either the analytic dense-plasma model of Lee and More \cite{lee-more-pof-1984}, specifically the GMS-6 model from Table I of Ref.\ \onlinecite{gericke-coullog-pre-2002}, or a more advanced LTE table if available.

\textbf{Mix models:} Reduced models of hydrodynamic mix due to Richtmyer-Meshkov (RM) and Rayleigh-Taylor (RT) instabilities are available.  These are designed to capture mix physics without spatially resolving it, and involve some parameters that must be tuned.  The methodology is detailed in Ref.\ \onlinecite{bachmann-mix-pop-2023}, which demonstrates good agreement between Lasnex mix modeling and experimental data for layered implosions with plastic capsules and separated nuclear reactants.  RT growth is calculated via a buoyancy-drag model \cite{zhou-buoyancy-drag-pre-2002}, which creates zones of mixed material at user-specified interfaces.  These mixed zones entail 3 subzones: one for each pure material, and a third that is atomically mixed.  We use an RM model that limits the growth to keep the perturbation ``finger tips'' between the shocked material interface and the transmitted shock \cite{hurricane-richtmyer-pof-2000}. Mix models are not used in the present work, but will be to understand the over-predicted yields on WarmMag gas-filled capsules, and for modeling magnetized ignition designs.

\textbf{Drive deficit: laser power multipliers:} LHT modeling differs in important ways from NIF ICF data. We discuss three discrepancies and physics modifications to reduce them.  The \textit{x-ray drive deficit} refers to the fact that the total hohlraum x-ray drive is higher in modeling than data.  This is seen in x-ray measurements with the DANTE detector, shock velocity measurements in ``keyhole'' shots with the VISAR detector, and in the time of peak capsule emission or ``bangtime'' \cite{jones-hohlraums-pop-2017}.  In this paper, we modify the laser power by a time-dependent factor (the same on all beams) to bring modeled shock velocity and bangtimes in line with measurements.  We also use an electron heat flux limiter $f$, which limits the magnitude of the electron heat flux $q$ to some fraction of the free-streaming value: $q=\min(fn_eT_ev_{Te}, \ka \nabla T_e)$. This is an old technique in HED modeling \cite{malone-fluxlim-prl-1975}, intended to mimic physical flux inhibition by things like nonlocal transport, microturbulence, or magnetic fields.  We follow the ``high flux model'' \cite{rosen-dca-hedp-2011} and use $f=0.15$ outside the capsule (as defined above), which generally has little effect on ICF hohlraums: similar results obtain with a larger value or the nonlocal Schurtz model \cite{schurtz-nonlocal-pop-2000}.  Inside the capsule we use $f=0.1$, which has little effect and is done to agree with high-resolution capsule modeling practice.  Jones et al.\ \cite{jones-hohlraums-pop-2017} revisited a low $f \approx 0.03$ to explain the drive deficit by allowing more unabsorbed inner-beam laser light to escape the opposite LEH as ``glinted light'' \cite{turnbull-glint-prl-2015}. Dedicated experiments to examine this show much less glint than $f=0.03$ modeling and generally support a high $f \approx 0.15$ \cite{lemos-glintraum-pop-2022, farmer-aps2023-pop-2024}. 

\textbf{Implosion hotspot shape: CBET clamp:} A second discrepancy with NIF data is with the compressed hotspot shape, as measured by x-ray or neutron images near peak emission.  Specifically the second Legendre moment or $P_2$ is a key quantity.  A major factor in $P_2$ is CBET between the inner and outer laser cones. This relies on accurate plasma conditions especially in the LEH.  This region has many overlapping laser beams and weakly collisional plasma.  Accurate modeling likely needs better accounting for the lasers and kinetic or other effects beyond standard rad-hydro.  LHT modeling generally produces too much CBET to be consistent with implosion shape data - assuming the non-CBET parts of the model that affect shape are correct. LLNL has for over 10 years used a time-dependent saturation clamp $\delta n_e/n_e$ on the amplitude of the CBET-driven ion acoustic waves.  Physical ion-wave nonlinearities can set in for $\delta n_e/n_e \sim 0.01$.  Values between 0.001 and 0.01 are typically needed to match shape data in hohlraums with low gas fill density $\leq$ 0.6 mg/cm$^3$. We stress that the $\delta n_e$ clamp at least partly stands in for incorrect plasma conditions instead of ion-wave nonlinearity.

\textbf{Outer-beam bubble conditions: Nernst multiplier:} The final discrepancy we discuss is with NIF measurements of flows and $T_e$ in the ``gold bubble'' plasma that expands from the hohlraum wall heated by the outer beams \cite{meezan-bubble-pop-2020}.  LHT modeling with a low flux limit $f=0.03$ is consistent with the data, while $f=0.15$ is not.  This is at odds with the inner-beam glint results.  It suggests a model with $f=0.15$ where the inners hit the wall but $f=0.03$ where the outers hit.  Recently, it was found that the MHD model with self-generated fields does this, if a Nernst multiplier $f_N=0.1$ is used to greatly reduce Nernst advection \cite{woods-nernst-privcomm-2023}. The Nernst effect advects $B$ field from hot to cold electrons.  In the hohlraum wall, this advects self-generated ``Biermann battery'' fields farther out in radius, to colder and denser plasma where the field is too small to magnetize the electron heat flow. The LHT adopts $f_N=0.1$ outside the capsule but $f_N=1$ (no modification) inside the capsule.

\section{Lasnex MHD Model}
We describe the MHD package in detail since it is a key part of this work.  Lasnex has had a ``toroidal'', or azimuthal, field package for four decades \cite{nielsen-mhdlasnex-1981}.  This includes a relatively complete implementation of Braginskii's classical fluid model \cite{braginskii-xport-1965}.  The ``poloidal'', or $rz$, MHD model is a more recent addition and does not yet implement all of Braginskii \cite{rambo-mhdlasnex-1998, rambo-vectorpot-lasnex-1998}.  Lasnex is an $rz$ code in that all quantities only vary in $r$ and $z$, but azimuthal $\phi$ vector components are of course allowed.  We write a vector $\myvec v = \myvec v_{rz} + v_\phi\hatphi$ with $\myvec v_{rz} = v_r\myhat r + v_z\myhat z$ the 2D vector in the $rz$ plane. The relevant fields are the magnetic field $\vecB$, vector potential $\myvec A$, electric field $\vecE$, net current $\vecJ$ and center-of-mass or bulk flow velocity $\vecu$.

We start with Maxwell's equations. $\vecB$ is written
\begin{align}
  \vecB     &= \vecB_{rz} + B_\phi \hatphi, \\
  \vecB_{rz} &= \nabla\times\vecA = \lp -\p_zA_\phi, r^{-1}\p_r(rA_\phi) \rp.
\end{align}
$\vecA = A_\phi\hatphi$ is only the vector potential for $\vecB_{rz}$, the full $\vecA$ would involve $A_r$ and $A_z$ for $B_\phi$.  $\vecB$ evolves in time by Faraday's law:
\begin{equation}
  \p_t\vecB = -\nabla\times\vecE.
\end{equation}
The $\phi$ component gives
\begin{equation} \label{eq:dBphidt}
  \p_tB_\phi = -(\nabla\times\vecE_{rz})\cdot\hatphi = \p_rE_z - \p_zE_r.
\end{equation}
The $r$ component gives $\p_z(\p_tA_\phi + E_\phi) = 0$, which we integrate over $z$ and set the boundary term to zero to find
\begin{equation}
  \p_tA_\phi = -E_\phi.
\end{equation}
The $z$ component gives the same result.  Faraday's law implies $\p_t(\nabla\cdot\vecB)=0$. We assume $\nabla\cdot\vecB=0$ initially so this remains true for all time, which gives
\begin{equation}
  \p_rB_r + \p_zB_z = -B_r/r.
\end{equation}
The current is given by Amp\`ere's law with no displacement current: $\mu_0\vecJ = \nabla\times\vecB$ which in components gives
\begin{align}
  \mu_0 \vecJ_{rz} &= \lp -\p_zB_\phi, r^{-1}\p_r(rB_\phi) \rp, \\
  \mu_0J_\phi &= \lp r^{-2} - r^{-1}\p_r - \p_{rr} - \p_{zz} \rp A_\phi.
\end{align}
$\vecE$ is given by Ohm's law below. It is convenient to label fields in the poloidal package as ``P'' fields: $\vecB_{rz}$, $A_\phi$, $E_\phi$ and $J_\phi$.  Similarly the toroidal or ``T'' fields are $B_\phi$, $\vecE_{rz}$ and $\vecJ_{rz}$.  We call a field ``P+T'' if it is nonzero only if both P and T fields are nonzero.  $\vecu$ is discussed below.

 The continuity  and bulk momentum equations are
 \begin{align}
   d_t\rho + \rho u_r/r + \rho(\p_ru_r + \p_zu_z) &= 0, \\
  \rho d_t \vecu_{rz} - \frac{\rho}{r} u_\phi^2 \myhat r + \nabla p + \nabla\cdot \myten\tau - \lp \vecJ \times \vecB \rp_{rz}  &= \vecF_{rz}, \\
  \rho d_t u_\phi + \frac{\rho}{r}u_ru_\phi - \lp \vecJ_{rz} \times \vecB_{rz} \rp_{\phi} &= F_\phi.
\end{align}
The gradient symbol $\nabla$ contains no $\phi$ derivative. These fields are summed over electron and ion species: mass density $\rho$, isotropic pressure $p$, anisotropic stress tensor $\myten\tau$, and other forces $\vecF$ not of current interest, from e.g.\ laser, radiation, and charged-particle deposition ($F_\phi$ is currently zero).  Terms with $u_\phi$ and no derivatives are centripetal forces and are only nonzero if $u_\phi$ is.  $\vecu_{rz}$ is always present and $(\vecJ \times \vecB)_{rz}$ is nonzero if \textit{either} P or T fields are present:
\begin{equation}
\lp \vecJ \times \vecB \rp_{rz} = \vecJ_{rz} \times (B_\phi \hatphi) + J_\phi \hatphi \times \vecB_{rz}.
\end{equation}
The only force that produces $u_\phi$ from an initial state with $u_\phi=0$ is
\begin{equation}
\lp \vecJ_{rz} \times \vecB_{rz} \rp_{\phi} = \mu_0^{-1}(B_z\p_zB_\phi + r^{-1}\p_r(rB_\phi)B_r).
\end{equation}
This force is a ``P+T'' field and is nonzero only if both P and T magnetic fields are nonzero. $d_t \equiv \p_t + u_r\p_r + u_z\p_z$ is the Lagrangian derivative in the $rz$ plane, but does not act on basis vectors: $d_t\vecu_{rz} = (d_tu_r) \myhat r + (d_tu_z)\myhat z$.  If one starts with $\vecu\cdot\nabla \vecu$ in Cartesian coordinates and transforms to cylindricals, one must account for spatial derivatives of the basis vectors $\myhat r$ and $\myhat z$. This gives rise to the centripetal terms.  There is no electric force since Lasnex assumes quasi-neutrality and thus the charge density $\sigma$ is zero.

We now discuss Ohm's law for $\vecE$, which is used to evolve $\vecB$.  Gauss's law $\nabla\cdot\vecE = \sigma/\epsilon_0$ is not used.  $\vecE$ is given by combining ion and electron momentum equations to find $\p_t\vecJ$ and setting it to zero. This is equivalent to setting $\p_t \vecu_e = 0$ in just the electron momentum equation. We also neglect advective terms $\vecu\cdot\nabla\vecu$ as higher order.  Using $m_e \ll m_i$, we find
\begin{align}
  \vecE &= -\vecu\times\vecB + \vecE_\text{nouB}, \\
  \vecE_\text{nouB} &=  (en_e)^{-1} \lp \vecJ\times\vecB -\nabla p_e + \vecF_{ep} + \vecR \rp.
\end{align}
We have separated the ideal-MHD term $\vecu\times\vecB$ and dropped the anisotropic electron stress tensor.  In $\vecE_\text{nouB}$, $\vecJ\times\vecB$ and $-\nabla p_e$ are the Hall and Biermann battery terms. $\vecF_{ep}$ is the laser ponderomotive force on electrons, which is the only force from the full $\vecF_e$ currently in Ohm's law. These 3 terms are fully implemented and not discussed further.  $\vecR$ is the e-i friction force and described below.  We start with Eq.\ \ref{eq:dBphidt} for $\p_tB_\phi$ and rewrite the ideal $\vecu\times\vecB$ term to find
\begin{equation}
  d_tB_\phi + \lp \p_ru_r + \p_zu_z \rp B_\phi - r\vecB_{rz}\cdot\nabla \frac{u_\phi}{r} = -(\nabla\times\vecE_{\text{nouB},rz})\cdot\hatphi.
\end{equation}
The first two terms on the left side reflect conservation of toroidal flux, or $B_\phi$ times the $rz$ area of a spatial zone.  The final left-side term is a source of $B_\phi$ due to twisting of the poloidal field by spatially varying azimuthal flow, or differential rotation. By similar logic, for the poloidal fields
\begin{equation}
  d_t \lp r A_\phi \rp = -\frac{r}{en_e} \lp \underbrace{\lp \vecJ_{rz} \times \vecB_{rz} \rp \cdot\hatphi} + R_\phi \rp.
\end{equation}
The poloidal Hall term with the underbrace is not currently implemented.

We now discuss the friction force $\vecR$, which is partly implemented.  We give the rationale for dropping various terms, though there is no rigorous justification. All terms are included for a purely T field. The full Braginskii form is
\begin{align}
  \vecR     &= (en_e)^{-1}\vecR_\al - \vecR_\be, \\
  \vecR_\al &= (\al_{||} - \al_\perp)(\hatb\cdot\vecJ)\hatb + \al_\perp\vecJ + \al_\wedge \vecJ\times\hatb, \\
  \vecR_\be &= (\be_{||} - \be_\perp)(\hatb\cdot\nabla T_e)\hatb + \be_\perp \nabla T_e + f_N\be_\wedge \hatb\times\nabla T_e. \label{eq:Rbeta}
\end{align}
$\hatb = \vecB/|\vecB|$, $\vecR_\al$ is the resistive force with different units from $\vecR$, and $\vecR_\be$ is the thermal force (note its sign). $f_N$ is the Nernst multiplier described above, and multiplies $\be_\wedge$ in all equations. Components of the collisional tensors $\al$, $\be$ and $\ka$ (see Eq.~(\ref{eq:qe}) below) are found from Sadler, Walsh and Li \cite{sadler-mhdcoef-prl-2021}, which is a corrected version of Epperlein and Haines \cite{epperlein-xport-pof-1986}. The components are functions of the effective charge state $Z_\text{eff} \equiv \sum_i Z_i^2 n_i / \sum_i Z_i n_i$ and the electron Hall parameter $\om_{ce}\tau_{eI}$ with $\om_{ce} \equiv eB/m_e$ and $\tau_{eI}$ the unmagnetized e-i collision time averaged over ion species (found as described above for e-i transport coefficients). Lasnex currently does not include P+T terms in $\vecR_\al$ which leaves
\begin{align}
  \vecR_{\al,rz} &= \al_\perp \vecJ_{rz} + \al_\wedge \lp \vecJ \times \hatb \rp_{rz}, \\
  R_{\al,\phi} &= \al_\perp J_\phi.
\end{align}
In $\vecR_\be$, Lasnex drops $\lp \hatb\cdot\nabla T_e \rp \hatb$: it is zero for a purely T field. This leaves
\begin{align}
  \vecR_{\be,rz} &= \be_\perp \nabla T_e + f_N\be_\wedge \lp \hatb\times\nabla T_e \rp_{rz}, \\
  R_{\be,\phi} &= f_N\be_\wedge \lp \hatb\times\nabla T_e\rp_\phi.
\end{align}

The electron energy density per volume $E_e$ obeys
\begin{align}
  d_tE_e + (E_e + p_e)\nabla\cdot\vecu + \nabla\cdot \vecq_e + \myten{\tau}_e : \nabla \vecu - \nabla \cdot \lp \vecJ E_e/en_e \rp \\
  - p_e\nabla\cdot(\vecJ/en_e) = (en_e)^{-1}\vecJ\cdot\vecR - \om_{ei}(T_e-T_i).  \nonumber
\end{align}
$\vecR$ includes the same terms as in Ohm's law, and $\om_{ei}$ is the e-i temperature equilibration rate. The electron heat flux $\vecq_e$ has units (speed * energy / volume) and is
\begin{align} \label{eq:qe}
  -\vecq_e &= \frac{T_e}{en_e} \lp \underbrace{(\be_{||}-\be_\perp)\vecJ\cdot\hatb \hatb_{rz}} + \be_\perp \vecJ_{rz} - f_N\be_\wedge (\vecJ\times\hatb)_{rz} \rp \\
  & +(\ka^e_{||}-\ka^e_\perp) (\hatb_{rz}\cdot\nabla T_e)\hatb_{rz} + \ka^e_\perp \nabla T_e + \ka^e_\wedge b_\phi\hatphi\times\nabla T_e. \nonumber
\end{align}
Lasnex does not currently include the term with the underbrace, which is P+T. $\ka^e$ is the electron thermal conductivity tensor, with units 1/(length*time). The ion energy equation is
\begin{equation}
    d_tE_i + (E_i + p_i)\nabla\cdot\vecu + \nabla\cdot \vecq_i + \myten{\tau}_i : \nabla \vecu = \om_{ei}(T_e-T_i).
\end{equation}
The ion heat flux is analogous to $\vecq_e$ with $\be=0$:
\begin{equation}
  -\vecq_i = (\ka^i_{||}-\ka^i_\perp) (\hatb_{rz}\cdot\nabla T_i)\hatb_{rz} + \ka^i_\perp \nabla T_i + \underbrace{\ka^i_\wedge b_\phi\hatphi\times\nabla T_i}.
\end{equation}
Lasnex does not currently include the term with the underbrace.  Usually $|\vecq_i| \ll |\vecq_e|$ but $\vecq_i$ can matter when magnetization strongly reduces $\vecq_e$.

\bibliography{c:/Users/strozzi2/latex/mybib}

\end{document}